\documentclass[preprint]{aastex62}
\usepackage{color}
\shorttitle{HCO$^{+}$ absorption line of NGC 1052}
\shortauthors{Sawada-Satoh et al.}
\begin{document}

\title{A broad HCO$^{+}$ absorption line associated with the circumnuclear torus of NGC 1052}

\correspondingauthor{Satoko Sawada-Satoh}
\email{swdsth@gmail.com}

%\author{%
%   Satoko Sawada-Satoh\altaffilmark{1,2}, 
%   Duk-Gyoo Roh\altaffilmark{3}, 
%   Se-Jin Oh\altaffilmark{3}, 
%   Sang-Sung Lee\altaffilmark{3,4}, 
%   Do-Young Byun\altaffilmark{3,4}, \\
%   Seiji Kameno\altaffilmark{5,6},
%   Jae-Hwan Yeom\altaffilmark{3},
%   Dong-Kyu Jung\altaffilmark{3},
%  Hyo-Ryoung Kim\altaffilmark{3},
%  Ju-Yeon Hwang\altaffilmark{3} 
%} 
    
%\altaffiltext{1}{Mizusawa VLBI Observatory, National Astronomical Observatory of Japan,
%2-12 Hoshigaoka-cho, Mizusawa-ku, Oshu, Iwate 023-0861 Japan; \myemaila}
%\altaffiltext{2}{Center for Astronomy, Ibaraki University, 2-1-1 Bunkyo, Mito, Ibaraki 310-8512, Japan; \myemailb}
%\altaffiltext{3}{Korea Astronomy and Space Science Institute, 
%776 Daedeok-daero, Yuseong, Daejeon 34055, Republic of Korea}
%\altaffiltext{4}{University of Science and Technology, 
%217 Gajeong-ro, Yuseong-gu, Daejeon 34113, Republic of Korea }
%\altaffiltext{5}{Joint ALMA Observatory, Alonso de Cordova 3107 Vitacura, Santiago 763 0355, Chile}
%\altaffiltext{6}{National Astronomical Observatory of Japan, 2-21-1 Osawa, Mitaka, Tokyo 181-8588, Japan}    

\author[0000-0001-7719-274X]{Satoko Sawada-Satoh}
\affil{Graduate School of Science and Engineering, Kagoshima University, 
1-21-35 Korimoto, Kagoshima 890-0065, Japan}

\author[0000-0003-1157-4109]{Do-Young Byun}
\affiliation{Korea Astronomy and Space Science Institute, 776 Daedeok-daero, Yuseong, Daejeon 34055, Republic of Korea}
\affiliation{University of Science and Technology, 217 Gajeong-ro, Yuseong-gu, Daejeon 34113, Republic of Korea}

\author[0000-0002-6269-594X]{Sang-Sung Lee}
\affiliation{Korea Astronomy and Space Science Institute, 776 Daedeok-daero, Yuseong, Daejeon 34055, Republic of Korea}
\affiliation{University of Science and Technology, 217 Gajeong-ro, Yuseong-gu, Daejeon 34113, Republic of Korea}

\author{Se-Jin Oh}
\affiliation{Korea Astronomy and Space Science Institute, 776 Daedeok-daero, Yuseong, Daejeon 34055, Republic of Korea}

\author{Duk-Gyoo Roh}
\affiliation{Korea Astronomy and Space Science Institute, 776 Daedeok-daero, Yuseong, Daejeon 34055, Republic of Korea}

\author[0000-0002-5158-0063]{Seiji Kameno}
\affiliation{Joint ALMA Observatory, Alonso de Cordova 3107 Vitacura, Santiago 763 0355, Chile}
\affiliation{National Astronomical Observatory of Japan, 2-21-1 Osawa, Mitaka, Tokyo 181-8588, Japan}

\author{Jae-Hwan Yeom}
\affiliation{Korea Astronomy and Space Science Institute, 776 Daedeok-daero, Yuseong, Daejeon 34055, Republic of Korea}

\author{Dong-Kyu Jung}
\affiliation{Korea Astronomy and Space Science Institute, 776 Daedeok-daero, Yuseong, Daejeon 34055, Republic of Korea}

\author{Chungsik Oh}
\affiliation{Korea Astronomy and Space Science Institute, 776 Daedeok-daero, Yuseong, Daejeon 34055, Republic of Korea}

\author{Hyo-Ryoung Kim}
\affiliation{Korea Astronomy and Space Science Institute, 776 Daedeok-daero, Yuseong, Daejeon 34055, Republic of Korea}

\author{Ju-Yeon Hwang}
\affiliation{Korea Astronomy and Space Science Institute, 776 Daedeok-daero, Yuseong, Daejeon 34055, Republic of Korea}
\affiliation{SET system, 16-3 Gangnam-daero 8-gil, Seocho-gu, Seoul 06787, Korea}

%\affiliation{AAS Director of Publishing}
%\affiliation{American Astronomical Society \\
%2000 Florida Ave., NW, Suite 300 \\
%Washington, DC 20009-1231, USA}

%\author{Jeff Lewandowski}
%\affiliation{IOP Senior Publisher for the AAS Journals}
%\affiliation{IOP Publishing, Washington, DC 20005}



\begin{abstract}

We present the first subparsec-scale maps of HCO$^{+}$ $J=$1--0 absorption 
in the circumnuclear region of the nearby radio galaxy NGC~1052. 
Our $\lambda$3mm VLBI observations with the Korean VLBI Network have spatially resolved 
the broad HCO$^{+}$ absorption at a velocity range of 1350--1850 km~s$^{-1}$ 
against a double-sided nuclear jet, 
and have revealed that the HCO$^{+}$ absorption is concentrated 
on the receding jet and the nuclear components. 
The distribution of the HCO$^{+}$ absorbing gas strongly supports the 
circumnuclear torus surrounding the supermassive black hole. 
%The column density of molecule hydrogen  is derived to be $10^{24}$--$10^{25}$  
%cm$^{-2}$, assuming LTE conditions and $T_{\rm ex}=$ 100--230 K. 
From the estimations of the column density and the volume density of molecular hydrogen, 
the size of the molecular gas region in the torus is at least 1~pc. 
The broad spectral profile of HCO$^{+}$ is likely to be a blend of 
multiple gas clumps with various velocities. 
The HCO$^{+}$ absorption of NGC~1052 could trace complex kinematics  
in the vicinity of the active galactic nucleus, 
such as inflow, outflow, turbulence, and so on. 

\end{abstract}

\keywords{galaxies: active --- galaxies: individual (NGC~1052)  --- galaxies: nuclei --- quasars: absorption lines}


\section{Introduction} \label{sec:intro}

NGC~1052 is a nearby radio galaxy with a systemic velocity 
($V_{\rm sys} = cz$) of 1507~km~s$^{-1}$ \citep{jensen03}. 
It exhibits a nearly symmetric double-sided radio jet 
from subparsec scales up to kiloparsecs 
\citep[e.g.,][]{jones84, wrobel84, kellermann98}, 
showing outward motions with an apparent velocity of 0.26$c$
\citep{vermeulen03}. 
Past very-long-baseline-interferometry (VLBI) studies have shown that 
the nuclear structure consists of 
the eastern approaching and the western receding jets 
at an inclination angle of $\ge57^{\circ}$ from the line of sight 
\citep[e.g.,][]{kellermann99, kameno01, kadler04b}. 
Atomic and molecular lines are found 
toward the center of NGC~1052 in emission 
(H$_2$O megamaser for \citealt{braatz94})
and in absorption 
(H~{\sc{i}} for \citealt{vangorkom86}; OH for \citealt{omar02}; 
HCO$^{+}$; HCN and CO for \citealt{liszt04})
at the radio band. 

NGC~1052 hosts a well-studied circumnuclear torus surrounding its 
central engine. 
The presence of a parsec-scale dense plasma torus has been proposed 
based on the measurements of free-free absorption in the innermost region 
of the radio jet obtained by the multi-frequency VLBI observations 
\citep{kameno01, kameno03, vermeulen03, kadler04b}. 
The torus is geometrically thick, obscuring 0.1~pc  and 0.7~pc 
of the eastern and western jets, respectively. 
The electron column density toward the free-free absorber is estimated 
to be $\sim10^{23}$ cm$^{-2}$, 
which is consistent with atomic hydrogen column density 
of $10^{22}$--$10^{24}$ cm$^{-2}$ 
derived from various past X-ray observations 
\citep{guainazzi99, weaver99, kadler04a}.  
\citet{kameno05} 
have proposed 
a torus model with several phase layers: 
a hot plasma layer at the inner surface,  
and a warm molecular gas layer where the H$_2$O maser arises, 
and  
a cooler molecular layer. 
Several spectral VLBI observations have revealed that 
gases of H$_2$O megamaser emission, OH absorption, and HCN absorption 
are located where the free-free absorption is large due to the torus, 
which supports the torus model with several layers  
\citep{sawadasatoh08, impellizzeri08, sawadasatoh16}. 
The spectra of the H$_2$O megamaser, OH absorption, and HCN absorption 
are all redshifted with respect to $V_{\rm sys}$, 
which has been explained as being due to ongoing material infalling 
onto the supermassive black hole
\citep{sawadasatoh08, impellizzeri08, sawadasatoh16}.
These circumnuclear structures make NGC~1052 an ideal laboratory to study 
subparsec-scale molecular chemistry and 
its relation to the  AGN-related physical and chemical process.

HCO$^{+}$ and HCN are known as good dense molecular gas tracers  
in galaxies, because of its large dipole moment and abundance. 
The IRAM Plateau de Bure Interferometer (PdBI) observations have discovered 
the HCO$^{+}$(1--0) absorption against NGC~1052 \citep{liszt04}. 
The derived spectral profile of the HCO$^{+}$ absorption was broad,  
extending from 1400 to 1900~km~s$^{-1}$ in velocity, 
and was not resolved into narrow absorption features. 
The peak velocity of the HCO$^{+}$ absorption was $\sim$1650~km~s$^{-1}$, 
which was $\sim$150~km~s$^{-1}$ redshifted from $V_{\rm sys}$. 
It was close to the peak velocities of HCN absorption \citep{liszt04} 
and H$_2$O maser emission \citep{braatz94}. 
To confirm what the broad 
HCO$^{+}$ absorption traces 
in the circumnuclear region of NGC~1052, 
one milliarcsecond (mas) angular resolution achieved by VLBI is essential. 
The VLBI map of molecular absorption line 
against the parsec-scale synchrotron radio source 
offers a unique scientific opportunity in direct detection of thermal molecular gas on parsec- and subparsec scales
 \citep[e.g.][]{sawadasatoh16}.

We have observed the high dense gas tracer HCO$^{+}$ transition 
toward the center of NGC~1052 with the 
Korean VLBI Network (KVN). 
Here we show the first detection and the maps of the subparsec-scale 
HCO$^{+}$(1--0) absorption 
in the circumnuclear region of NGC~1052.  
One mas corresponds to 0.095 pc in the galaxy.

\section{Observations and data reduction} \label{sec:obs}

KVN observations of NGC~1052 were carried out 
from 2017 June 17 UT 21:00 to June 18 UT 06:00, 
for a total on-source time of 7.5 hours. 
To improve the sensitivity 
for the $\lambda$3mm HCO$^+$(1--0) absorption line observation,
simultaneous dual-frequency observation was conducted 
at $\lambda$1.3cm and $\lambda$3mm bands 
using the KVN multi-frequency receiving system \citep{han08, oh11, han13}.
To cover the broad HCO$^+$(1--0) absorption feature 
($\sim$400~km~s$^{-1}$),  
the data were recorded at each station with the Mark6 system 
at a sampling rate of 8~Gbps (512~MHz $\times$ 4~IFs $\times$ 2~bit quantization) 
in dual circular polarization. 

Two of four IFs were assigned to left-hand circular polarization (LHCP) and right-hand circular polarization (RHCP) 
at $\lambda$3mm 
for the target frequency band. 
The other two IFs were fixed to LHCP and RHCP 
at $\lambda$1.3cm  
for the phase referencing. 
The velocity coverage of one IF 
at $\lambda$3mm 
was $\sim$1600~km~s$^{-1}$. 
For phase and bandpass calibration, 
3C~84 and NRAO~150 were also observed every hour.

The data were correlated with the DiFX software correlator \citep{deller07}
at the Korea-Japan Correlation Center \citep{yeom09, lee15b}.
The visibility amplitude decrement due to the digital quantization loss 
was corrected by applying a correction factor of 1.1 \citep{lee15a}. 
Post-correlation processing was done using NRAO AIPS software 
\citep{greisen03}.
A priori amplitude calibration was applied using measurements 
of the opacity-corrected system temperature 
with the chopper-wheel method \citep[e.g.][]{ulich76} 
and gain curve depending on the elevation. 
Complex bandpass characteristics at each station were solved 
using  data of 3C~84 and NRAO~150.  
%The cross-power spectra of them were then used to estimate the phase responses of the bandpass filters. 

To calibrate rapid atmospheric phase fluctuations 
at $\lambda$ 3mm, 
we analyzed these data using the frequency phase transfer (FPT) method, 
in which the phase solutions of the low-frequency band 
($\lambda$1.3cm) 
are transferred 
to the high-frequency band 
($\lambda$3mm) 
by scaling by their frequency ratio 
\citep{middelberg05, rioja11}. 
Applying FPT, 
the rms phase fluctuation was reduced by $\sim40\%$ for 
all of three baselines. 
Doppler velocity corrections were made by running the AIPS tasks SETJY and CVEL.
The continuum was subtracted with the AIPS task UVLSF 
by performing a polynomial fit to line-free channels in the visibility domain. 
We identified channels 
with a velocity lower than 1400~km~s$^{-1}$ and higher than 1900~km~s$^{-1}$
as the line-free channels. 
The parallel-hand data were averaged into the total intensity data. 
We corrected the visibility phase using self-calibration with the averaged line-free channels, 
and applied the solutions of the self-calibration in the absorption line channels.
The continuum map was formed from the averaged line-free channels
using the hybrid imaging with CLEAN and self-calibration. 
The image cube was generated with a channel width of 8~MHz. 
The optical depth image cube was yielded from 
the continuum map and the image cube, and averaged every 32~MHz. 
We blanked image pixels with intensities below 18~mJy~beam$^{-1}$ ($< 6\sigma$)
in the continuum map, 
because the signal-to-noise ratio in optical depth is poor where the continuum emission 
is weak. 
Maps were produced with natural weighting, 
and the resulting FWHM of the synthesized beam is $1.60\times0.87$ mas 
($0.152\times0.082$ pc in NGC 1052).

\section{Results} \label{sec:results}

\begin{figure}[ht!]
%\plotone{fig1_rev.eps}
\begin{center}
    \includegraphics[scale=0.85]{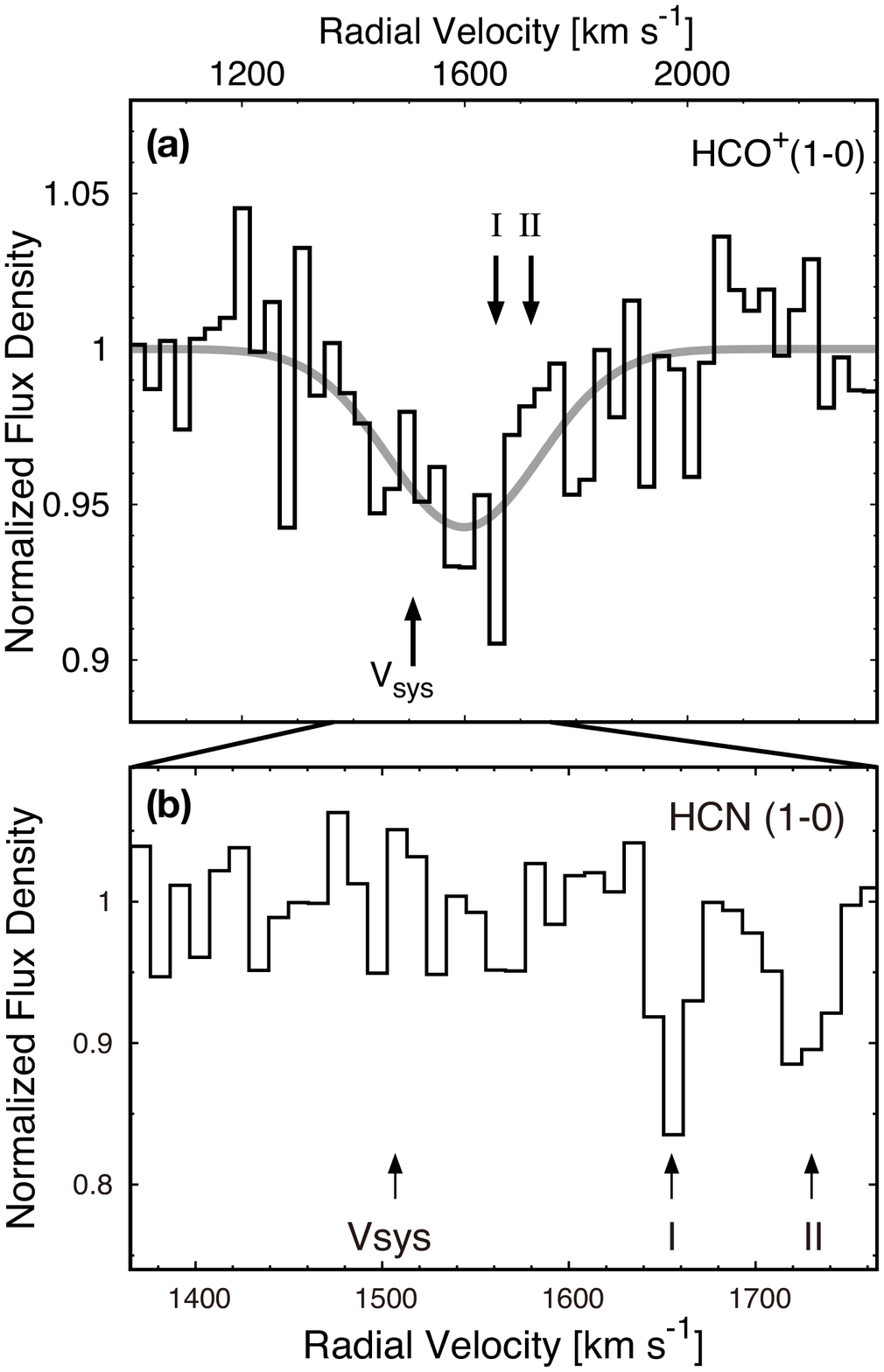}
\end{center}
\caption{
(a) Cross power spectrum of HCO$^{+}$(1--0) absorption lines 
integrated over the whole on-source time of  NGC~1052 obtained with the KVN, 
represented with the black solid line. 
The velocity resolution is 26.9 km~s$^{-1}$, 
and the rms noise in the normalized flux density is 0.02. 
The gray solid line gives a Gaussian fit to the absorption profile.  
The labels I and II represent the narrow HCN absorption features 
at 1656 and 1719~km~s$^{-1}$, detected with the past KVN observation 
\citep{sawadasatoh16}.  
(b) Spectral profile of HCN (1--0) absorption of NGC~1052 obtained with the KVN 
in 2015. The velocity resolution is 10.5 km~s$^{-1}$. 
This panel is exactly the same as Figure 1 in \cite{sawadasatoh16}. 
\label{spc}
}
\end{figure}

Figure~\ref{spc}a shows the vector-averaged cross power spectrum of 
HCO$^{+}$(1--0) absorption of NGC~1052 
integrated 
over the whole on-source time with the KVN. 
The frequency resolution is 8~MHz, which corresponds to 26.9~km~s$^{-1}$ in velocity resolution ($\Delta \nu$). 
All the spectral channels are normalized to the continuum level. 
Broad HCO$^{+}$(1--0) absorption is detected in the velocity range 
from 1350 to 1850 km~s$^{-1}$. 
The absorption profile could be slightly asymmetrical with the blueshifted part 
having more extended wing than the redshifted part, 
and the peak absorption channel is at velocity of 1658~km~s$^{-1}$. 
The peak channel is 
close in velocity to the narrow HCN absorption features I at 1656~km~s$^{-1}$, 
reported in  
\cite{sawadasatoh16} (figure~\ref{spc}b). 
The HCO$^{+}$ spectral profile, 
including a redshifted peak and a blueshifted wing, 
and their velocities 
are similar to those of the absorption profile from the PdBI data \citep{liszt04}. 
However, 
the peak absorption channel at 1658 km~s$^{-1}$ has 
a depth of $-9.5 \%$ of the continuum level,  
which is deeper than that of the PdBI ($-3\%$). 
The peak absorption depth results in the peak optical depth ($\tau_{\rm p}$) of 
$0.10\pm0.02$. 

To give a rough estimation of the width and the maximum depth, 
we attempted a single Gaussian fitting to the spectral profile, 
while it is rather asymmetrical. 
We obtained the best-fit model profile with
the full widths at half depth in velocity of 
$272\pm50$ km~s$^{-1}$
and the maximum depth of $-0.06\pm0.01$   
 at the centroid velocity of $1599\pm25$ km~s$^{-1}$.  
The reduced chi square is 1.04, and it is not further improved even if we give 
multiple-gaussian fitting. 

%%%%%%%%%%%%

\begin{figure*}
   \plotone{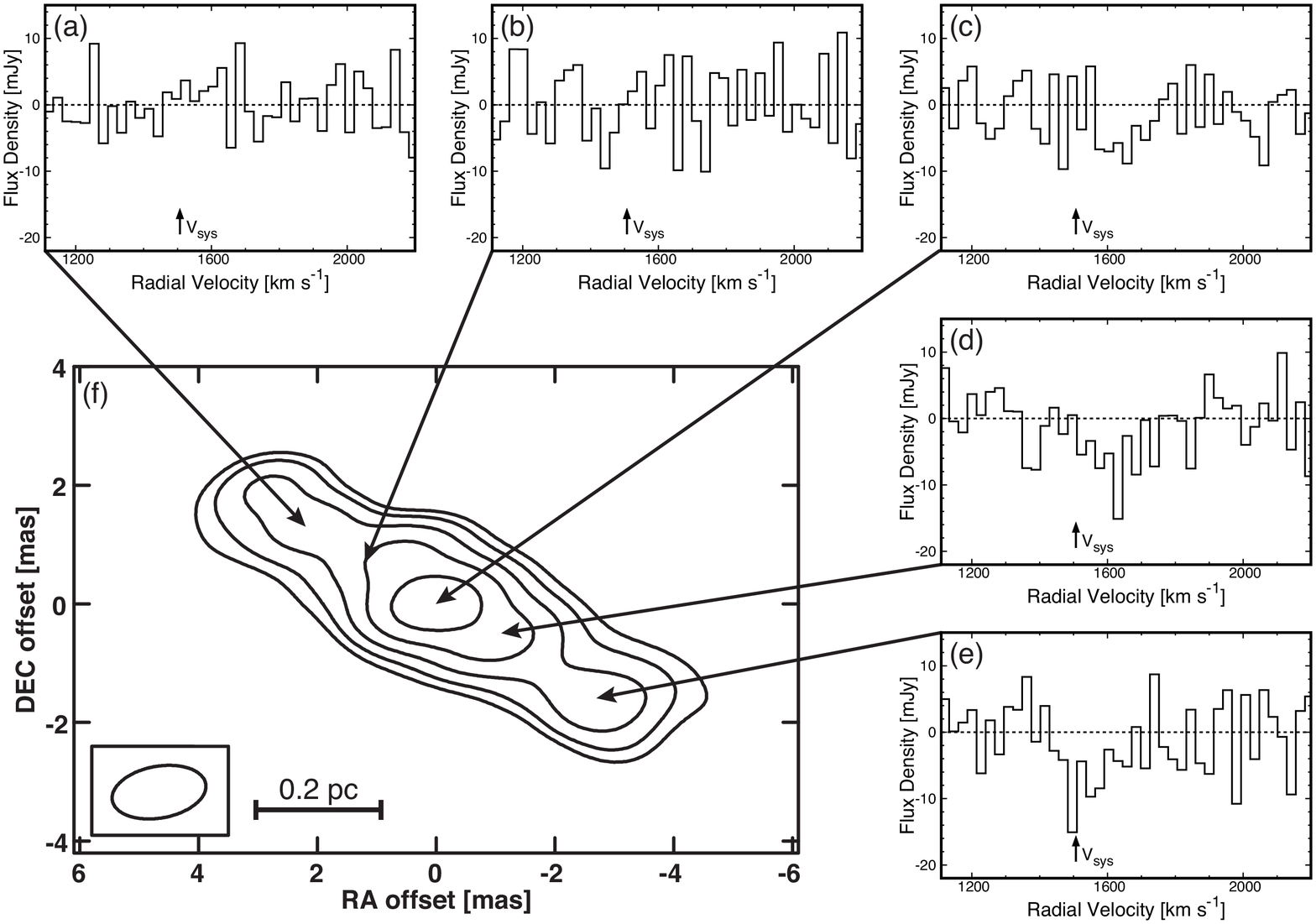}
 \caption{(a)--(e) Spectra of HCO$^{+}$ (1--0) absorption 
 at different locations as marked in the arrows. 
 The velocity resolution is 26.9 km~s$^{-1}$. 
 Zero flux density corresponds to the continuum level for each location. 
(f) Continuum image of the nuclear region in NGC~1052 at $\lambda$3mm. 
The contour starts at the $3 \sigma$ level, increasing by a factor of 2, 
where $\sigma=3$~mJy~beam$^{-1}$. 
The peak intensity of the brightest component is 244 mJy~beam$^{-1}$. 
The synthesized beam size is shown at the lower-left corner. 
 }\label{mapspc}
\end{figure*}

The broad HCO$^{+}$ absorption is spatially resolved by the VLBI imaging. 
In figure~\ref{mapspc}, we present a continuum image 
of the nuclear region in NGC~1052 
at $\lambda$3mm 
and 
HCO$^{+}$(1--0) absorption spectra in various regions  
against the background continuum emission of NGC~1052. 
The HCO$^{+}$(1--0) absorption spectra  
were extracted from the image cube using the AIPS task ISPEC 
integrating over the region of $1.4\times1.4$ mas 
at different spatial positions. 
The typical 1$\sigma$ level is 5~mJy in 8~MHz (26.9 km~s$^{-1}$) channel. 
Figures~\ref{mapspc}a and \ref{mapspc}b indicate that 
no significant absorption feature is seen 
against the eastern approaching jet,  
while a weak absorption feature is tentatively detected on 
the central nuclear component at velocity of $\sim$1650~km~s$^{-1}$ 
as shown in figure~\ref{mapspc}c. 
Figures~\ref{mapspc}d and \ref{mapspc}e show that 
the absorption depth reaches  3$\sigma$ level 
on the western receding jet. 

The $\lambda$3mm continuum image reveals 
a symmetric double-sided jet structure 
that consists of a bright central component and two elongated 
eastern and western jets 
(figure~\ref{mapspc}f). 
The jet structure is extended up to $\sim$0.7~pc. 
The flux densities of the central component within 1.6~mas 
(the major axis of the synthesized beam), 
the eastern jet, and the western jet 
are determined to be 414, 92, and 93~mJy, 
by summing the CLEAN components. 
The continuum image is consistent with the previous KVN map at 89~GHz 
taken in 2015 \citep{sawadasatoh16}. 

\begin{figure*}
  \plotone{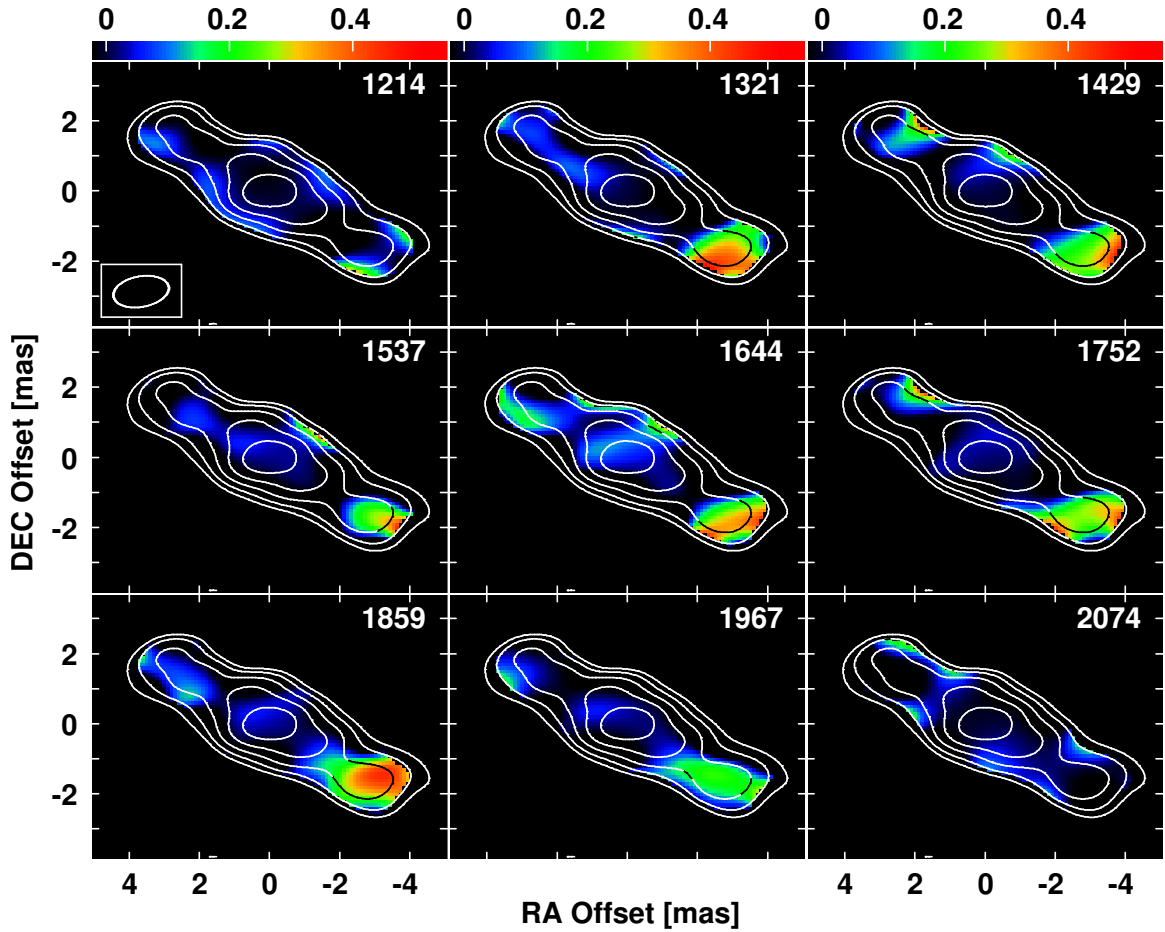}
 \caption{Color channel maps of the HCO$^{+}$(1--0) optical depth 
 averaging every 32~MHz (108 km~s$^{-1}$), 
 overlaid by a contour map of $\lambda$3mm continuum emission 
 (Figure~\ref{mapspc}f). 
 Color indicates the optical depth of HCO$^{+}$ absorption. 
 Their central velocities are shown at the upper right. 
 The rms noise in optical depth is 0.02 and 0.09 
 on the central and western receding jet components,
 respectively.   
 }\label{chmap}
\end{figure*}

HCO$^{+}$ optical depth channel maps superimposed 
on the $\lambda$3mm continuum image 
are shown in figure~\ref{chmap}. 
Distribution of HCO$^{+}$ optical depth on the double-sided jet structure shows that 
high opacities are localized on the western receding jet side, where 
the parsec-scale torus obscures. 
Such a localization of the high opacity on the receding jet side is also seen in 
the opacity distributions of OH \citep{impellizzeri08} and HCN \citep{sawadasatoh16} 
of this galaxy.

\section{Discussions} \label{sec:discussions}

%\subsection{Covering factor}

From our KVN observations, 
the HCO$^{+}$ opacity of $\sim$0.4 is derived in the nuclear region of NGC~1052 
with the subparsec angular resolution. 
It is one order higher than that obtained from the PdBI 
observations \citep{liszt04}. 
This fact suggests that 
the HCO$^{+}$ covering factor is 
much larger on subparsec scales.  
Here we examine it 
by estimating the mean opacity over the whole $\sim$0.7~pc nuclear structure 
for each frequency channel 
$\langle \tau_{\nu} \rangle$, using 
\begin{equation}
\langle \tau_{\nu} \rangle = 
\frac{\int \!\! \int \tau_{\nu} (x,y) \: I(x,y) \: dx \: dy}
{\int \!\! \int I(x,y) \: dx \: dy}, 
\end{equation}
where $\tau_{\nu} (x,y)$ is the HCO$^{+}$ opacity distribution for each channel map (color map in figure~\ref{chmap})
and $I(x, y)$ is the $\lambda$3mm continuum image (figure~\ref{mapspc}f). 
We also calculated normalized flux density for each channel from $\langle \tau_{\nu} \rangle$, and listed it in table~\ref{tab:meanopa}.
The derived normalized flux density is in the range from 0.96 to 1, 
and it is in good agreement with the HCO$^{+}$ absorption profile on hundred-parsec scales 
observed with the PdBI \citep{liszt04}. 
Thus, the HCO$^{+}$ covering factor varies on scales in the center of NGC~1052, 
and 
the difference in HCO$^{+}$ optical depth between 
KVN and PdBI could account for 
the partial covering of HCO$^{+}$ absorbing gas on subparsec scales.

% ===  Table sample === %
\begin{deluxetable}{ccc}
\tablecaption{Mean Opacity over the Whole Nuclear Source for Each Channel}
\tablehead{
\colhead{Central Velocity} &  \colhead{$\langle \tau_{\nu} \rangle$} & \colhead{Normalized Flux Density}  \\
\colhead{(km s$^{-1}$)}    &  \colhead{}  & \colhead{} 
%\colhead{(1)} & \colhead{(2)} & \colhead{(3)} 
 }
 \colnumbers
 \startdata
1214 & 0.0014 & 0.999 \\
1321 & 0.0191 & 0.981 \\
1429 & 0.0165 & 0.984 \\
1537 & 0.0038 & 0.996 \\
1644 & 0.0384 & 0.962 \\
1752 & 0.0198 & 0.980 \\
1859 & 0.0400 & 0.961 \\
1967 & 0.0137 & 0.986 \\
2074 & 0.0040 & 0.996 \\
\enddata
\tablecomments{ 
(1) Central velocity for each channel;  
(2) mean opacity over the whole parsec-scale nuclear source; 
(3) normalized flux density calculated from the values in (2). 
}
\end{deluxetable} \label{tab:meanopa}

%\subsection{The relation to the circumnuclear torus}

Concentration of high HCO$^{+}$ opacity on the western receding jet component 
strongly suggests that HCO$^{+}$ absorption is associated 
with the parsec-scale circumnuclear torus. 
Because the jet axis of NGC~1052 is oriented 
within $\sim30^{\circ}$ from the sky plane 
\citep[e.g., ][]{kellermann99, kameno01, kadler04b}, 
the near side of the thick torus should lie in front of the nuclear component and the receding jet component.  
The line of sight to the receding jet passes through the near side of the torus, 
where the HCO$^{+}$ absorption occurs. 
The HCO$^{+}$ gas presents a relatively common distribution 
with plasma, H$_2$O, OH, and HCN.  
If the torus has several physical layers as shown 
in figure~3 of \cite{sawadasatoh16}, 
the HCO$^{+}$ gas could be located in the cooler ($<$400~K) molecular layer 
inside the torus, where the HCN gas also could lie.

%\subsection{Molecular column density }

We can estimate the total column density of HCO$^{+}$ 
assuming the local thermodynamic equilibrium, 
using the approximate expression for the $J=$ 1--0 transition 
\begin{equation}
N_{\rm tot}=\frac{3k (T_{\rm ex}+hB/3k)}
{8 \pi^3 B \mu^2 \lbrack 1-\exp{(-h\nu/kT_{\rm ex})}\rbrack }
\int \tau dv ,
\end{equation}
where $k$ is the Boltzmann constant, $h$ is the Planck constant, 
$\mu$ is the permanent dipole moment of the molecule, 
$B$ is the rotational constant, $T_{\rm ex}$ is the excitation temperature, 
and $\int \tau d\nu$ is the velocity-integrated optical depth of the absorption feature.
For HCO$^{+}$, $\mu$ is 4.07~Debye \citep{haese79} and 
$B=44594$ MHz  \citep{lattanzi07}. 
As $T_{\rm ex}$ is uncertain, here we assume $T_{\rm ex}=$ 100 and 230~K, 
in the same manner as \citet{sawadasatoh16}. 
Assuming $\int \tau dv = \tau_{\rm p} \Delta \nu$ for the peak absorption channel 
at 1658 km~s$^{-1}$, 
the total column density of HCO$^{+}$ of the peak is 
calculated to be ($1.5\pm0.3$)$\times10^{15}$ cm$^{-2}$ 
and ($7.6\pm1.5$)$\times10^{15}$ cm$^{-2}$ at 100 and 230~K, respectively. 
The column density ratio between HCN and HCO$^{+}$ around 1658~km~s$^{-1}$ 
can be determined to be $\sim$6.5, 
as the total HCN column density at 1656~km~s$^{-1}$ 
is $9.5\times10^{15}$ cm$^{-2}$ 
and $5.0\times10^{16}$ cm$^{-2}$ at 100 and 230~K, respectively \citep{sawadasatoh16}. 
The derived value of the column density ratio is even higher compared to 
the high HCN/HCO$^{+}$ intensity ratios  
($R_{\rm HCN/HCO^{+}} \sim$ 2) 
measured in NGC~1097 \citep{izumi13} and NGC~1068 \citep{garcia14, viti14} 
on parsec scales with the ALMA. 
However, we have to note that the column densities of HCN and HCO$^{-1}$ 
are not measured simultaneously. 
During a time gap of 27 months between 
the two observations of HCN and HCO$^{+}$, 
the background receding jet component could move 0.18~pc outward, 
almost one beam size. 
Adopting an abundance ratio HCO$^{+}$ relative to H$_2$ of (2--3) $\times10^{-9}$ 
in Galactic diffuse molecular gas \citep{liszt00, liszt04, liszt10}, 
the column density of molecular hydrogen (H$_2$) is derived to be 
$N_{{\rm H}_{2}} \sim 10^{24}$--$10^{25}$ cm$^{-2}$, 
which is consistent with $N_{{\rm H}_{2}}$ estimated 
from the HCN absorption spectrum  
\citep{sawadasatoh16}. 

% critical density

Despite the detection of HCO$^{+}$ (1--0) absorption,  
the emission of HCO$^{+}$ has not been found in the center of NGC~1052 
with the past PdBI observation \citep{liszt04}. 
This indicates that the H$_2$ volume density ($n_{{\rm H}_2}$) in the center is less than 
the H$_2$ critical density ($n_{\rm cr}$) for HCO$^{+}$ (1--0) emission.  
Adopting the Einstein coefficient for spontaneous emission 
$A_{10} = 4.52\times10^{-5}$ s$^{-1}$ for HCO$^{+}$ \citep{izumi13}, and 
the collisional rate 
$\gamma_{10}$ of $1.8\times10^{-10}$ cm$^{-3}$ s$^{-1}$ at a temperature of 100~K \citep{flower99}, 
gives a critical density 
$n_{\rm cr} = A_{10}/\gamma_{10}$ of approximately $2.5\times10^5$ cm$^{-3}$. 
Thus, it derives $n_{{\rm H}_2} <  2.5\times10^5$ cm$^{-3}$. 
The relation $N_{{\rm H}_{2}}=n_{{\rm H}_2} f_v L$, 
where $f_v$ is the volume-filling factor and $L$ is the size of the molecular gas region in the torus, 
gives a lower limit $L > 1/f_v$ pc, 
using $N_{{\rm H}_{2}} = 10^{24}$ cm$^{-2}$ and 
$n_{{\rm H}_2} <  2.5\times10^5$ cm$^{-3}$ at 100~K. 
If we assume that the molecular region is inhomogeneous (i.e. $f_v < 1$), 
$L$ would be even larger than 1~pc.

It is remarkable that 
a broad HCO$^{+}$ absorption with a redshifted peak and a blueshifted wing 
is detected against the parsec-scale receding jet of NGC~1052. 
The broad spectral profile is 
much wider than expected from the thermal broadening ($\sim$0.2~km~s$^{-1}$), 
and it suggests a significant contribution of other kinematics, or
a complex of some rapid motion with multiple clumpy gas clouds at various different velocities, such as turbulence, interaction, and so on. 
The spectral profile was not resolved into several narrow absorption features. 
However, the optical depth differential among different velocity channel maps 
implies that several gas clumps or inhomogeneous structure 
at various velocities lie along the line of sight and apparently overlap. 
It is likely that 
the redshifted HCO$^{+}$ absorption around the peak velocity 
trace the same infall motion 
as HCN absorption and H$_2$O maser inside the torus. 
It can be interpreted as infall toward the supermassive black hole of NGC~1052. 
In addition to the infall motion, 
the blueshifted HCO$^{+}$ absorbing gas could indicate small-scale turbulence of clumps 
inside the torus. 

\section{Summary}

Our 1 mas angular resolution observations 
toward the center of NGC~1052 with the KVN 
have led us to first subparsec-scale imaging 
of the HCO$^{+}$ absorbing gas 
in the vicinity of AGN. 
The spectral profile of HCO$^{+}$ absorption is detected 
in a broad velocity range of 1350--1850~km~s$^{-1}$, 
in agreement with the past PdBI observations. 
However, the peak absorption depth is deeper than that of the PdBI. 
Our HCO$^{+}$ optical channel map clearly shows  
a high opacity on the receding jet component and a faint opacity on the nuclear component.  
It suggests that HCO$^{+}$ absorption arises from the near side of the torus, 
which covers the receding jet and the nucleus. 
We estimate $N_{{\rm H}_2}$ of $10^{24}$--$10^{25}$ cm$^{-2}$, assuming an abundance ratio 
[HCO$^+$]/[H$_2$] of 2--3 $\times 10^{-9}$ and a $T_{\rm ex}$ of 100--230~K. 
Since HCO$^+$ line appears only in absorption 
even with the PdBI, 
we find $n_{\rm H_2} < n_{\rm cr} = 2.5\times10^5$ cm$^{-3}$. 
It implies that the radius of the torus is $>$1~pc. 
The HCO$^{+}$ absorbing is probably a several gas clumps with several different velocities, 
rather than a single uniform medium with a single velocity. 
The broad HCO$^{+}$ absorption could consist of several kinematics at various velocities 
inside the torus.  

\acknowledgments

We are grateful to all members of KVN for supporting our observations.
The KVN and a high-performance computing cluster are facilities operated 
by the KASI (Korea Astronomy and Space Science Institute). 
The KVN observations and correlations are supported through the high-speed network connections 
among the KVN sites provided by the KREONET (Korea Research Environment Open NETwork), 
which is managed and operated by the KISTI (Korea Institute of Science and Technology Information).
SK was supported by JSPS KAKENHI Grant Number JP18K03712. 
SSL was supported by the National Research Foundation of Korea (NRF) grant funded 
by the Korea government (MSIP) (NRF-2016R1C1B2006697).


\vspace{5mm}
\facilities{KVN}

% ===  Figure 2 === %

% ===  Figure 3 === %

\end{document}